\documentclass[%
 aip,
 jap,
 amsmath,amssymb,
 reprint,%
]{revtex4-1}

\usepackage{graphicx}
\usepackage{dcolumn}
\usepackage{bm}

\usepackage[utf8]{inputenc}
\usepackage[T1]{fontenc}
\usepackage{mathptmx}
\usepackage{etoolbox}

\usepackage[colorlinks=true,
            linkcolor=black,
            urlcolor=black,
            citecolor=black]{hyperref}
\usepackage{tikz}

\usepackage{upgreek}
\newcommand\copyrighttext{%
  This paper has been accepted for publication in the \textit{Journal of Applied Physics}. Personal use of this material is permitted. However, permission from AIP Publishing must be obtained for all other uses, including reproduction, redistribution, or reuse of any portion of this work in other works, as well as for advertising or promotional purposes, or for creating new collective works for resale or redistribution.}

\newcommand\copyrightnotice{%
\begin{tikzpicture}[remember picture,overlay]
\node[anchor=south,yshift=10pt] at (current page.south) {\fbox{\parbox{\dimexpr0.85\textwidth-\fboxsep-\fboxrule\relax}{\copyrighttext}}};
\end{tikzpicture}%
}
\renewcommand\fbox{\fcolorbox{red}{white}}
\makeatletter
\def\@email#1#2{%
 \endgroup
 \patchcmd{\titleblock@produce}
  {\frontmatter@RRAPformat}
  {\frontmatter@RRAPformat{\produce@RRAP{*#1\href{mailto:#2}{#2}}}\frontmatter@RRAPformat}
  {}{}
}%
\makeatother
\begin{document}
\preprint{AIP/123-QED}

\title{A mobile high spatial-resolution Muography instrument based on large-area Micromegas detectors}
\author{Yu Wang}
\affiliation{State Key Laboratory of Particle Detection and Electronics, University of Science and Technology of China, Hefei 230026, China}
\affiliation{Department of Modern Physics, University of Science and Technology of China, Hefei 230026, China}

\author{Shubin Liu*}
\email{liushb@ustc.edu.cn}
\affiliation{State Key Laboratory of Particle Detection and Electronics, University of Science and Technology of China, Hefei 230026, China}
\affiliation{Department of Modern Physics, University of Science and Technology of China, Hefei 230026, China}
\affiliation{School of Nuclear Science and Technology, University of Science and Technology of China, Hefei 230026, China}

\author{Zhihang Yao}
\affiliation{State Key Laboratory of Particle Detection and Electronics, University of Science and Technology of China, Hefei 230026, China}
\affiliation{School of Nuclear Science and Technology, University of Science and Technology of China, Hefei 230026, China}

\author{Yulin Liu}
\affiliation{State Key Laboratory of Particle Detection and Electronics, University of Science and Technology of China, Hefei 230026, China}
\affiliation{Department of Modern Physics, University of Science and Technology of China, Hefei 230026, China}

\author{Zhiyong Zhang}
\affiliation{State Key Laboratory of Particle Detection and Electronics, University of Science and Technology of China, Hefei 230026, China}
\affiliation{Department of Modern Physics, University of Science and Technology of China, Hefei 230026, China}

\author{Zhengyang He}
\affiliation{State Key Laboratory of Particle Detection and Electronics, University of Science and Technology of China, Hefei 230026, China}
\affiliation{Department of Modern Physics, University of Science and Technology of China, Hefei 230026, China}

\author{Ziwen Pan}
\affiliation{State Key Laboratory of Particle Detection and Electronics, University of Science and Technology of China, Hefei 230026, China}
\affiliation{Department of Modern Physics, University of Science and Technology of China, Hefei 230026, China}

\author{Changqing Feng}
\affiliation{State Key Laboratory of Particle Detection and Electronics, University of Science and Technology of China, Hefei 230026, China}
\affiliation{Department of Modern Physics, University of Science and Technology of China, Hefei 230026, China}

\date{\today}

\begin{abstract}
Muon radiography is an imaging technique based on muon absorption in matter that allows measurement of internal details in hidden objects or structures.
This technique relies on measuring cosmic-ray muons tracks accurately, which reflects the incoming muon flux from both the target object and the open sky.
In this paper, we report on the construction of a high spatial resolution muography instrument based on Micromegas detectors.
Using four layers of $\mathrm{400~mm\times 400~mm}$ Micromegas detectors, channel multiplexing circuits, and the versatile readout system, a moveable muography instrument named $\upmu$STC-R400 was designed and constructed.
Results show that the channel multiplexing circuits can resolve hit positions correctly, and the spatial resolution of the detector is approximately $190~\upmu\mathrm{m}$.
Experiments were conducted at an under-construction subway tunnel and outdoors near a mountain, demonstrating the $\upmu$STC-R400's ability to maintain high spatial resolution outside the laboratory and its robustness in harsh environments.
\end{abstract}

\maketitle
\copyrightnotice
\section{Introduction}
Cosmic-ray muon, with its strong penetrating power and naturally occurring, is an ideal probe for imaging the internal structure of large or well-shielded objects.
Currently, muon imaging (muography) primarily employs two principles: muon transmission radiography and muon scattering tomography. 
This paper primarily focuses on muon transmission radiography, a technique based on the attenuation of muon flux as they pass through large structures, such as geological formations or man-made architectures. 
When a muon traverses a material, it experiences energy loss primarily through ionization and radiative processes. 
The mean energy loss rate can be described by the Bethe–Bloch and Bremsstrahlung equations. 
For a thin layer $dl$ of material with density $\rho (l)$, the energy loss primarily depends on the mass thickness, $\rho \times dl$.
The fraction of muons crossing a material is therefore determined by the integrated density over the path length, as shown in Eq.(\ref{eq_opacity}), a quantity called opacity, also known as mass thickness.~\cite{procureurMuonImagingPrinciples2018}

\begin{equation}
 x(L)=\int_{L} \rho(l) d l=\bar{\rho}\times L 
 \label{eq_opacity}
 \end{equation} 

\begin{figure}[htbp]
\centerline{\includegraphics[width=0.5\textwidth]{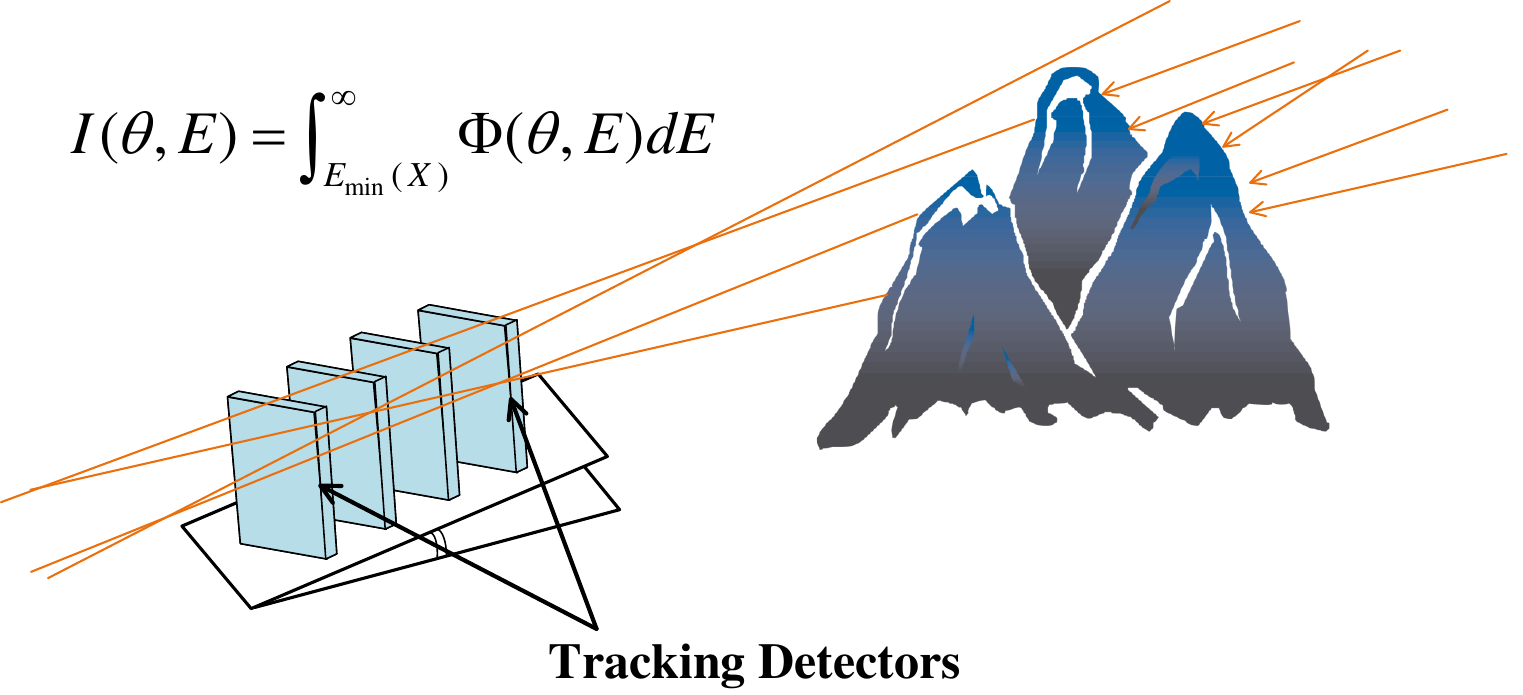}}
\caption{
Principle of muon radiography. 
A tracking system, typically an array of position-sensitive detectors, measures the incident muon flux. 
As muons traverse an object, their flux is attenuated due to absorption and scattering.
}
\label{fig1}
\end{figure}

The principle of the experimental measurement of the fraction of muons crossing a material is illustrated in Fig.~\ref{fig1}, where the muon flux is measured by tracking detectors both toward the object ($N_{\text{object}}(\theta, \phi)$) and the open sky ($N_{\text{open-sky}}(\theta, \phi)$). 
By calculating the ratio $N_{\text{object}}/N_{\text{open-sky}}$, the transmission ratio of cosmic-ray muons can be determined, which is expressed as follows:
\begin{equation}
 T(\theta, \phi)=\frac{N_{\text {object }}(\theta, \phi)}{N_{\text {open-sky }}(\theta, \phi)}=\frac{\int_{E_{\min }(X)}^{\infty} \Phi(\theta, E) d E}{\int_{E_{0}}^{\infty} \Phi(\theta, E) d E} 
 \label{eq_transmission}
\end{equation}
Here, $\Phi(\theta, E)$ is the differential muon flux, $E_{\min}(x)$ is the minimum energy required for a muon to traverse an object with opacity $x$, and $E_0$ represents the minimum energy threshold for muons detected (e.g., determined by detector characteristics or analysis cuts). 
In experimental analysis, the differential muon flux $\Phi(\theta, E)$ is primarily dependent on the zenith angle $\theta$. 
Therefore, the incident flux $\Phi(\theta, E)$ is considered to be the same for both the open-sky and object measurements within the same angular bin.

This transmission ratio is directly linked to the energy loss experienced by muons, which in turn depends on the material's opacity. 
Consequently, for a muon to traverse a material of a given opacity $x(L)$, it must possess a minimum initial energy, $E_{min}$. 
Once $E_{min}$ is determined, the opacity $x$ can be calculated. Subsequently, the average density $\bar{\rho}$ can be obtained if the physical length $L$ of the object is known.

To convert the measured transmission ratio $T$ into opacity $x$ (and subsequently density), several approaches are commonly employed to model the muon spectrum. 
These include comprehensive Monte Carlo simulations of cosmic-ray air showers (such as those performed with CORSIKA (COsmic Ray Simulations for KAscade)~\cite{gottowikCORSIKA8Modern2025}) which can generate detailed muon energy and angular distributions.
Empirical models, derived from experimental data, are also used; For instance, the EcoMug library~\cite{paganoEcoMugEfficientCOsmic2021} is based on measurements from ADAMO detectors~\cite{bonechiDevelopmentADAMODetector2005}. 
Additionally, some methods utilize analytical calculations of pion and kaon decay to generate the energy spectrum of cosmic-ray muons, with notable examples including models proposed by Gaisser et al.~\cite{gaisserCosmicRaysParticle1991} and Bugaev et al.~\cite{bugaevAtmosphericMuonFlux1998}.
Hebbeker et al. further refined these by combining Bugaev's spectrum with experimental data~\cite{hebbekerCompilationHighEnergy2002}.
Alternatively, for simpler analyses or specific experimental conditions, the energy loss rate can be approximated as a constant. 
For instance, in the DIAPHANE experiment~\cite{marteauDIAPHANEMuonTomography2017}, the muon energy loss in rock was set at $\mathrm{2.5~MeV\cdot cm^2/g}$.

The first application of muon radiography was the measurement of the overburden mass of the Guthega-Munyang tunnel using a Geiger counter array as the sensitive detectors~\cite{GEORGE1955}. 
Subsequently, in the 1960s, L.W. Alvarez explored the possibility of searching for hidden chambers within the Second Pyramid of Giza~\cite{l_w_alvarez_search_1970}.
Over the past three decades, numerous experiments have been conducted in areas such as volcano imaging~\cite{kazahayaDegassingProcessSatsumaIwojima2002,tanakaDetectingMassChange2009,bonechiMURAVESProjectOther2018,loprestiMuographicMonitoringVolcanotectonic2020}, geological surveying~\cite{liuDeepInvestigationMuography2024,nishiyamaFirstMeasurementIcebedrock2017a,nishiyamaBedrockSculptingActive2019}, and detection of cavities for archaeological~\cite{beniUseMuonImaging2025} and civil engineering applications~\cite{jourdeMonitoringTemporalOpacity2016a,scampoliCosmicRayMuography2023}. 
A notable recent example is the discovery of a big void in Khufu's Pyramid with the combination of nuclear emulsion chambers, plastic scintillator detectors, and Micromegas detectors~\cite{Kunihiro_2017}.

The key aspect of muography is the accurate measurement of muon tracks. 
Except the nuclear emulsion chambers, most detection methods employ multiple layers of detectors to reconstruct these tracks.
The use of finer resolution detectors can obtain more accurate results and contribute to a more compact design.
Spatial resolution and detection area are two critical parameters for optimizing imaging time and accuracy.
Imaging time is determined by the number of muons that penetrate the object and are subsequently recorded by the detectors. 
A muography instrument with a larger detector area has a greater acceptance, allowing the detection of muons from a wider range of directions, which helps to maintain a reasonable experimental duration. 

The most common detector type employed in muography is the scintillator with Silicon Photomultiplier (SiPM) readout~\cite{niculescu-oglinzanuSiROScintillatorbasedHodoscope2024,baccaniMIMAProjectDesign2018,marteauDIAPHANEMuonTomography2017,anastasioMURAYExperimentApplication2013,derrico_MURAVESMuonTelescope_2022}. 
A typical scintillator telescope consists of two or more double layers of orthogonal plastic scintillator bars, used to measure muons tracks in two independent projection directions. 
For rectangular bars, where a single muon typically produces a hit in one scintillator bar, the spatial resolution $\sigma$ is related to the bar's lateral width $L$ by $\sigma =L/\sqrt{12}$, assuming a uniform distribution of hit positions within the bar~\cite{bonechi_AtmosphericMuonsImaging_2020}. 
For triangular-shaped bars, spatial resolution can be improved by measuring the signal fraction in adjacent bars, achieving approximately $\mathrm{3~mm}$ resolution for a bar with $\mathrm{1.7~cm}$ height and $\mathrm{3.3~cm}$ width~\cite{derrico_MURAVESMuonTelescope_2022}.

With advancements in nuclear instrumentation, micro-pattern gaseous detectors (MPGDs) have become a feasible option for muon radiography applications, as demonstrated by their uses in imaging Khufu's Pyramid~\cite{Kunihiro_2017} and WatTo experiment~\cite{bouteilleMicromegasbasedTelescopeMuon2016}. 
These detectors can achieve a spatial resolution of a few hundred micrometers at a reasonable cost. 
In particular, the Micromegas (Micro-MEsh Gaseous Structure) detector can achieve a spatial resolution of approximately 100 $\upmu \mathrm{m}$ over active areas of up to several thousands square centimeters. 
However, a large detection area combined with fine spatial resolution increases the complexity of the readout system design. 
This challenge can be mitigated through specially designed channel multiplexing methods, which reduce the number of required readout channels while preserving spatial information, achieving a balance between detection area and readout complexity~\cite{procureurGeneticMultiplexingFirst2013,qiNovelMethodEncoded2016,bouteilleLargeResistive2D2016,yuan_2DEncodedMultiplexing_2017,wangHighCompressionRatioChannelMultiplexingMethod2025}.
On the other hand, gaseous detectors are often considered less suitable for field applications outside the laboratory due to their requirement for a continuous gas supply and their susceptibility to discharge caused by environmental factors such as vapor or dust pollution. 
These issues can be addressed by carefully controlling the gas flow and by robustly enclosing the detectors.

In this paper, a high spatial-resolution muon radiography instrument based on thermal-bonding Micromegas detectors is developed and presented.
The sensitive area of this detector is $\mathrm{400~mm \times 400~mm}$, and the number of anode strips is 2000 for a single detector with two dimension readout.
To achieve a compact instrument design, a novel channel multiplexing method and highly integrated front-end electronics cards have been implemented in this system.
This detector achieves a spatial resolution better than $200~ \upmu \mathrm{m}$ with the implemented multiplexing circuit and front-end electronics. 
Subsequently, we performed muon radiography experiments in a subway tunnel under construction, and an outdoor environment near a mountain, demonstrating the capability and stability of this instrument.

\section{SYSTEM DESIGN AND EXPERIMENTAL SETUP}
The muography instrument consists of four layers of Micromegas detectors serving as sensitive elements, channel multiplexing circuits to reduce readout electronics requirements, Front-end Electronics Cards (FECs) for signal amplification and digitization, and back-end electronics for event identification and data concentration.

\subsection{Resistive Micromegas Detector}
\begin{figure}[htbp]
\centerline{\includegraphics[width=0.5\textwidth]{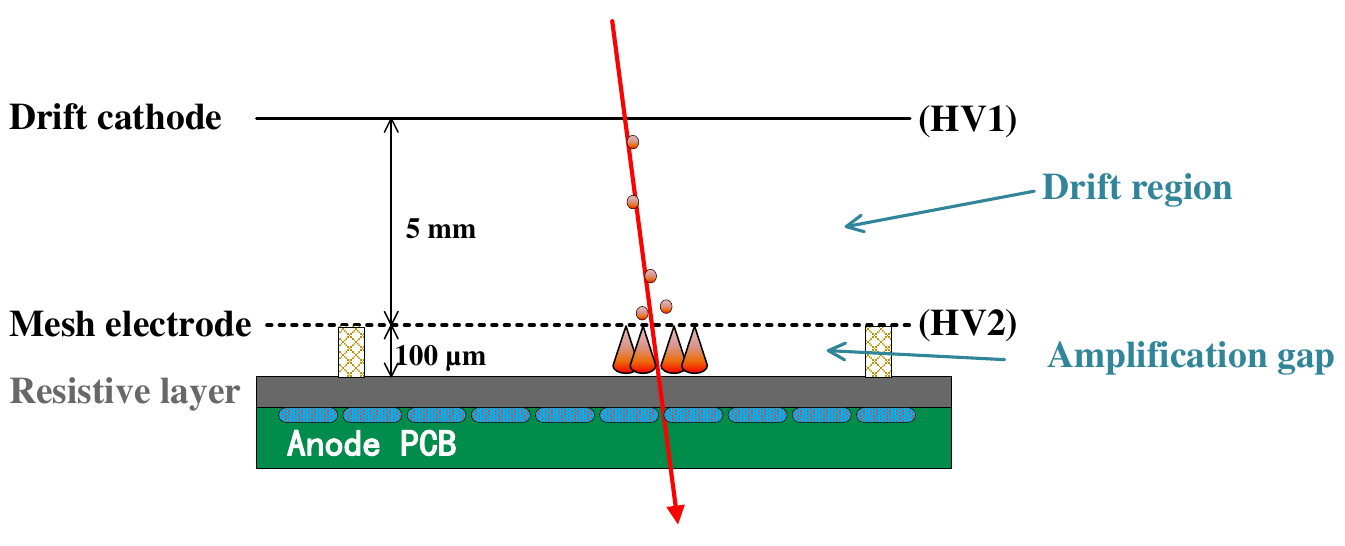}}
\caption{Simplified cross-sectional structure of the Micromegas detector. Note that the thicknesses of the drift region and the amplification gap are not drawn to scale.
}
\label{fig2}
\end{figure}

The Micromegas detector is a typical micro-pattern gaseous detector operating in proportional mode, featuring a drift region and a thin amplification gap structure. 
A simplified cross-sectional structure of the Micromegas is shown in Fig.~\ref{fig2}. 
The active volume is separated into a drift region, where incident particle ionizes the working gas, and an amplification gap, where the primary ionizations are multiplied through the avalanche process.

The distance between the drift cathode and mesh electrode of our detector is $\mathrm{5~mm}$, and the thickness of the amplification gap is approximately $100~\upmu \mathrm{m}$.
A germanium layer coated on the surface of the anode PCB (Printed Circuit Board) serves as a resistive anode and is connected to the GND. 
The conversion-drift electric field is established by applying negative voltages to the mesh electrode and a slightly higher voltage to the drift cathode .

When a muon enters the detector, primary ionization is generated. 
The strong electric field in the amplification region draws the primary electrons through the mesh voids and initiates the avalanche process. 
The signal induced on the anode strips is a sum of the electrons and ions signal, and this signal, when read out by a low-noise charge-sensitive preamplifier, is primarily due to the positive ion drift towards the micromesh electrode.

\begin{figure}[htbp]
\centerline{\includegraphics[width=0.5\textwidth]{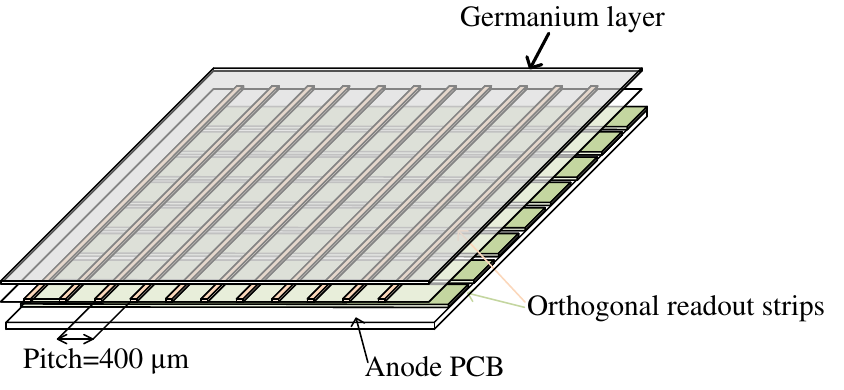}}
\caption{Structure of the readout strips on the anode PCB, where the second and third layers form the orthogonal readout strips. The germanium layer is coated after PCB manufacture.
}
\label{fig3}
\end{figure}

As shown in Fig.~\ref{fig3}, the readout strips were designed in the second and third layers of the PCB with an orthogonal arrangement. 
This design makes a single active volume to measure both X and Y hit positions. 
The strip pitch is $400~\upmu\mathrm{m}$, with the width of strips in the upper layer being $68~\upmu\mathrm{m}$ and those in the lower layer being $295~\upmu\mathrm{m}$.

In contrast to conventional Micromegas fabrication methods, such as the bulk method based on photoetching processes, we employed a thermal bonding method.
In this approach, small spacers made of thermal bonding film are placed on the anode PCB to support the tensile mesh electrode.
Further details about the Micromegas detector can be found in this work~\cite{jianxinfengThermalBondingMethod2021}, where the detector design was enlarged for the present work.

\begin{figure}[htbp]
\centerline{\includegraphics[width=0.5\textwidth]{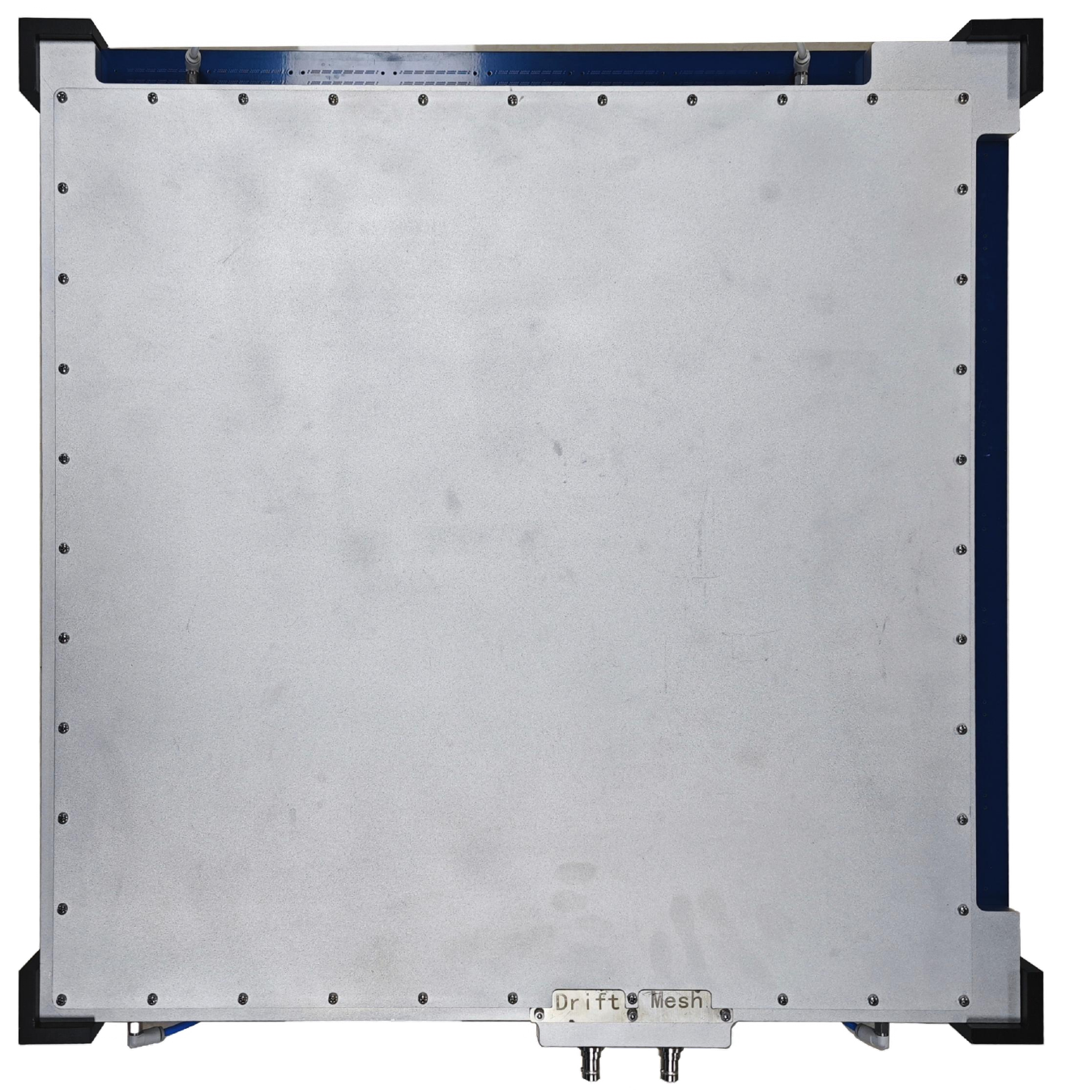}}
\caption{Photograph of the Micromegas detector ($\mathrm{400~mm\times 400~mm}$) used in this research.
}
\label{fig4}
\end{figure}

Several detectors with different dimensions were designed and implemented.
In this paper, detectors with sensitive areas of $\mathrm{400~mm\times 400~mm}$ are used to construct the muon radiography instrument.
A photograph of the Micromegas detector is shown in Fig.~\ref{fig4}.

\subsection{Channel multiplexing Circuit}
A major constraint in the application of high-spatial resolution Micromegas detectors is the requirement for a large number of readout channels, leading to increased cost, power consumption, and data processing complexity.
For example, the Micromegas detector used in this work has 2000 anode strips (1000 for X and 1000 for Y dimensions), which would typically require 2000 readout channels.
For a muon radiography instrument with four layers of detectors, the total number of strips requiring readout could reach 8,000.
To address this challenge, we proposed a position-encoded multiplexing method, in which each front-end electronics channel reads out multiple detector strips.

The fundamental principle of this multiplexing approach is that within a single acquisition window (typically set to a few microseconds or less), at most one muon will hit the detector, which is critical to avoid ambiguities in hit reconstruction.
The resulting ionization charges distribute over several adjacent strips for each dimension, near the hit point.
Therefore, a specially designed mapping relation can be implemented, where each contiguous group of strips of the detector in the same layer is connected to a unique pair of readout electronics channels. 
When signals are read back on a specific group of multiplexed readout channels, multiple potential hit locations on the detector could be inferred.
However, according to the mapping relationship, only one result corresponds to contiguous strips, which is the actual hit location on the detector.
A mathematical model for this approach was established in our previous work, where different types of multiplexing circuits were designed and evaluated~\cite{wangHighCompressionRatioChannelMultiplexingMethod2025}.

\begin{figure}[htbp]
\centerline{\includegraphics[width=0.5\textwidth]{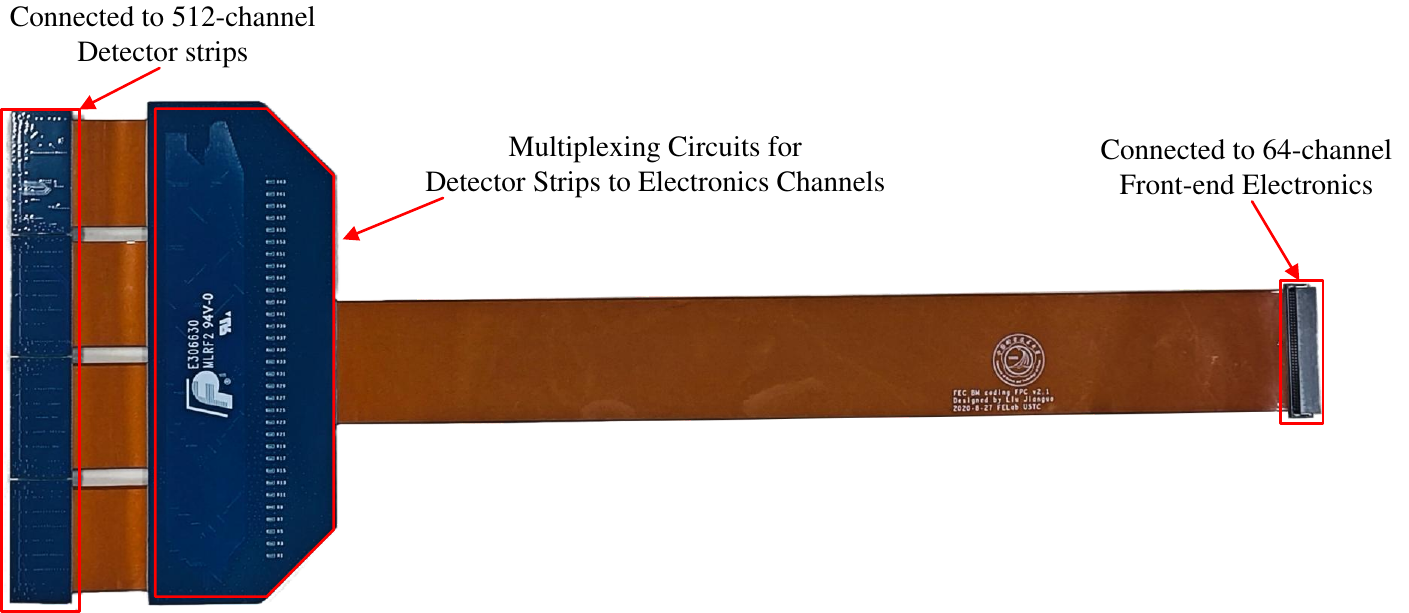}}
\caption{
Photograph of the multiplexing circuit that connects 512 detector strips to 64 readout electronics channels.
}
\label{fig5}
\end{figure}

In this work, we employ a circuit with a multiplexing factor of eight, where 512 detector strips are multiplexed into 64 front-end electronics channels.
The encoding circuit is shown in Fig.~\ref{fig5}. 
For each dimension of the detector strips, two multiplexing circuits are used, meaning that 1024 strips of the detector are read out by 128 channels of electronics.

\subsection{Front-end Electronics Card (FEC)}
The readout is handled by an FEC which adopts a design that was verified in the PandaX III (Particle AND Astrophysical Xenon experiment III) project~\cite{d.zhuDevelopmentFrontEndElectronics2019}.
The key differences lie in the use of an upgraded version of the readout ASIC named STAGE (SEcond stAGe of the AGET~\cite{anvarAGETGETFrontend2011,baudinASTREASICSwitched2018a}) and the utilization of channel hit information for self-trigger signal generation.

\begin{figure}[htbp]
\centerline{\includegraphics[width=0.5\textwidth]{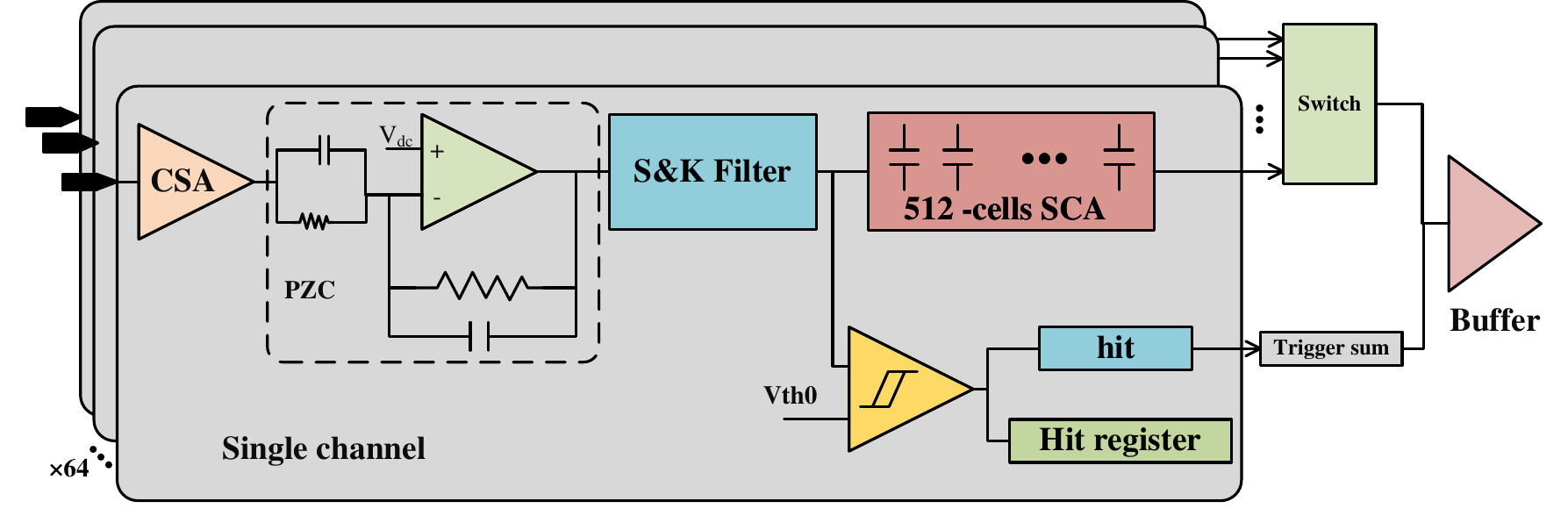}}
\caption{
Block diagram of the STAGE chip.
}
\label{fig6}
\end{figure}

The structure of the STAGE chip is shown in Fig.~\ref{fig6}. 
It has 64 active channels which process the signals from the Micromegas detector after channel multiplexing, and 4 dummy channels which are not connected to the detector but serve as noise monitors for the chip. 
Each channel consists of a charge sensitive amplifier (CSA), a pole-zero canceller (PZC), an analog shaper ( S\&K Filter,  Sallen–Key Filter), a discriminator for trigger generation, and a 512-cell switched capacitor array (SCA).
The amplified and shaped signal is sampled by the SCA and then sent out of the chip under the control of the field-programmable gate array (FPGA) on the FEC. 
The trigger signal from each channel contributes to the total trigger signal, with the amplitude proportional to the number of channels exceeding the on-chip threshold.

A single FEC uses four STAGE chips to sample the encoded signals, with one FEC reading out exactly one layer of the detector. 
The output of each STAGE chip is digitized by a single-channel 12-bit analog-to-digital converter (ADC) and processed by the FPGA.
A channel-specific threshold is set to suppress the idle channels. 
All digitized data with values greater than the preset thresholds, along with timestamp, trigger number, and checksum, are constructed into a user-defined data frame and transmitted to the back-end data acquisition (DAQ) board.

\subsection{Back-end Electronics and Remote Control}
A back-end electronics card designed for small to medium-scale physics experiments is adopted in the construction of this muography instrument~\cite{liuBackendElectronicsBased2025}. 
It performs data gathering from the FECs, event identification of potential muons, and data transmission to the host PC.

The back-end electronics card communicates with the FECs via optical fiber interfaces using a custom-defined protocol.
With the time domain multiplexing (TDM) method, the link between the back-end electronics and FECs is divided into three virtual channels for trigger, data, and command transmission.

For remote control, a 4G modem provides Internet access to the instrument. 
A Windows Subsystem for Linux (WSL) based automatic data processing procedure performs data processing and transmission to the laboratory server.
Manual control, remote debugging, and testing are performed via Windows Remote Desktop through a ZeroTier network.

\subsection{Self-trigger mode}
A self-trigger mode was developed to select potential muon signals while the instrument operates in continuous mode.
This mode consists two main parts: pre-trigger generation by the FECs, which indicates the detection of over-threshold signals, and subsequent valid trigger generation by the back-end electronics after performing trigger selection.

The pre-trigger signal is generated through two stages in the FEC. 
The first stage is implemented using the built-in `multiplicity' function of the ASIC~\cite{baudinASTREASICSwitched2018a}.
As shown in Fig.~\ref{fig6}, during the acquisition phase, the shaped signal is continuously sampled by the SCA and simultaneously compared with a programmable voltage threshold.
If the signal exceeds the threshold, a single-channel hit signal is generated. 
A total trigger signal is then produced based on the single-channel hit signals in an analog mode, with an amplitude proportional to the number of single-channel hits.
This total trigger signal (referred to as the multiplicity signal, which indicates how many channels are over threshold) is amplified to the differential analog output ports and continuously acquired by the on-board ADC. 
A second-stage trigger is generated by applying two thresholds to identify valid multiplicity signals: a lower one to suppress triggers caused by noise, and a higher one to eliminate spurious triggers resulting from discharges in the Micromegas detectors.
When the multiplicity signal falls between the lower threshold and the higher threshold, a potential event is detected, and the FEC sends a pre-trigger signal to the back-end electronics.

The back-end electronics records pre-trigger signals from each FEC and maintains them for a programmable time period, after which the register is cleared.
If the recorded pre-triggers match the configured trigger pattern, a fast trigger signal followed by a 32-bit trigger ID is distributed to all FECs.
The trigger pattern is configurable via control commands. 
In this specific application, a valid trigger is generated and sent to the FECs if at least three layers of FECs generate pre-trigger signal.
Upon receiving the fast trigger signal, the FECs stop the acquisition phase and controls the ASICs to sequentially output the sampled analog signals from the SCAs for conversion.
The digitized data is compared against the preset channel-specific thresholds, and only data exceeding these thresholds is packaged and transmitted to the back-end electronics.

\subsection{Construction of the Instrument}

\begin{figure}[htbp]
\centerline{\includegraphics[width=0.45\textwidth]{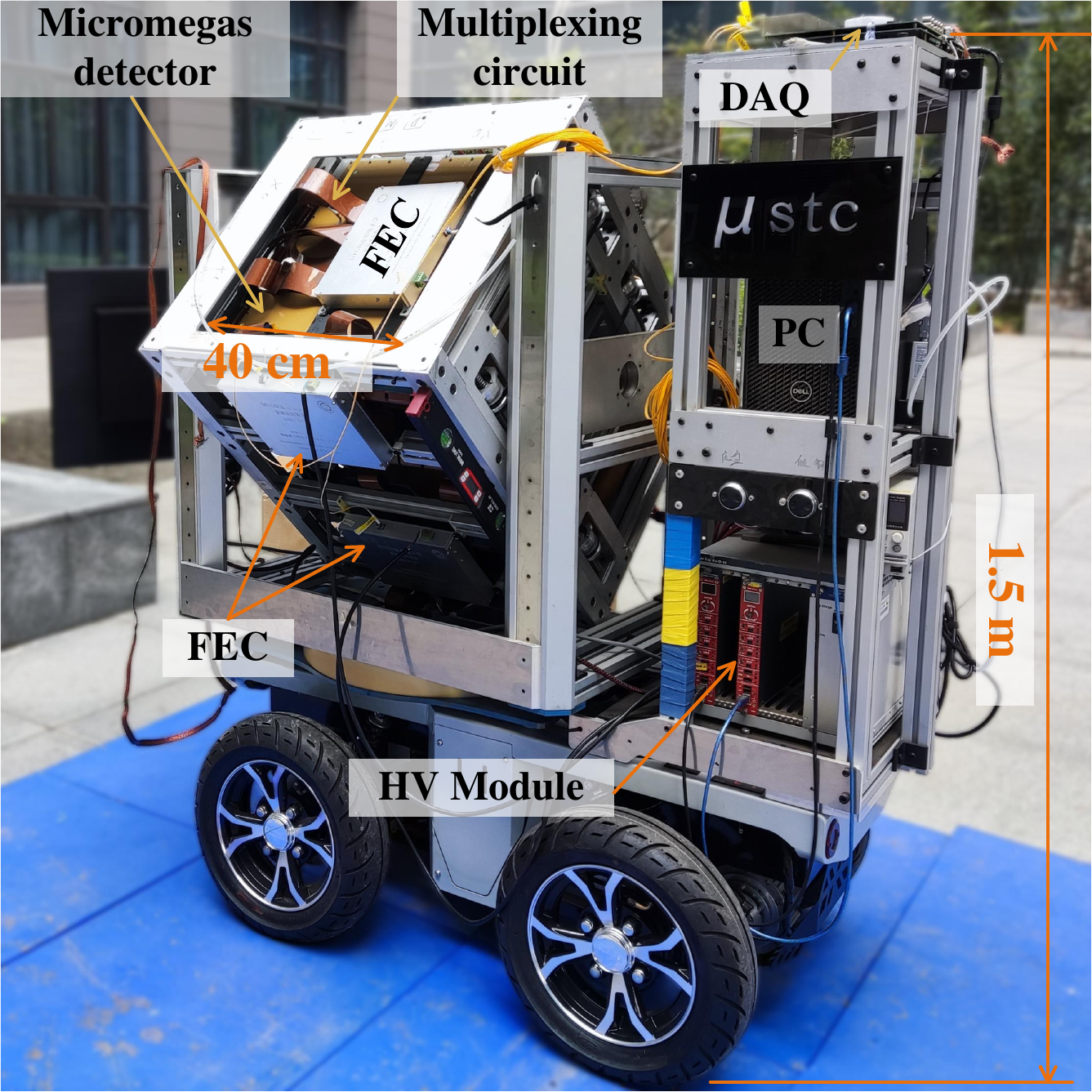}}
\caption{
Photograph of the $\upmu$STC-R400. The constructed system includes Micromegas detectors, multiplexing circuit, FECs, back-end electronics (DAQ) card, PC, HV modules, and the four-wheel chassis.
}
\label{fig7}
\end{figure} 
A muon radiography instrument was constructed using these detectors and electronics cards, and named $\upmu$STC-R400 (muon Scattering and Transmission imaging faCility, utilizing $\mathrm{400~mm\times 400~mm}$ detectors for radiography). 
As shown in Fig.~\ref{fig7}, this radiography instrument contains four layers of Micromegas detectors, sixteen multiplexing circuits (four for each layer), four FECs, a back-end electronics card, a mini-PC, two high-voltage (HV) modules housed in a mini-NIM case, and a remote-controlled Ackerman chassis to support the entire system.
The system's orientation, including its zenith and azimuthal angles, is controlled by two motors, allowing for precise adjustment of the viewing direction towards target objects.

\section{Experimental Results}
\subsection{Performance of the Readout Electronics}
\begin{figure}[htbp]
\centerline{\includegraphics[width=0.45\textwidth]{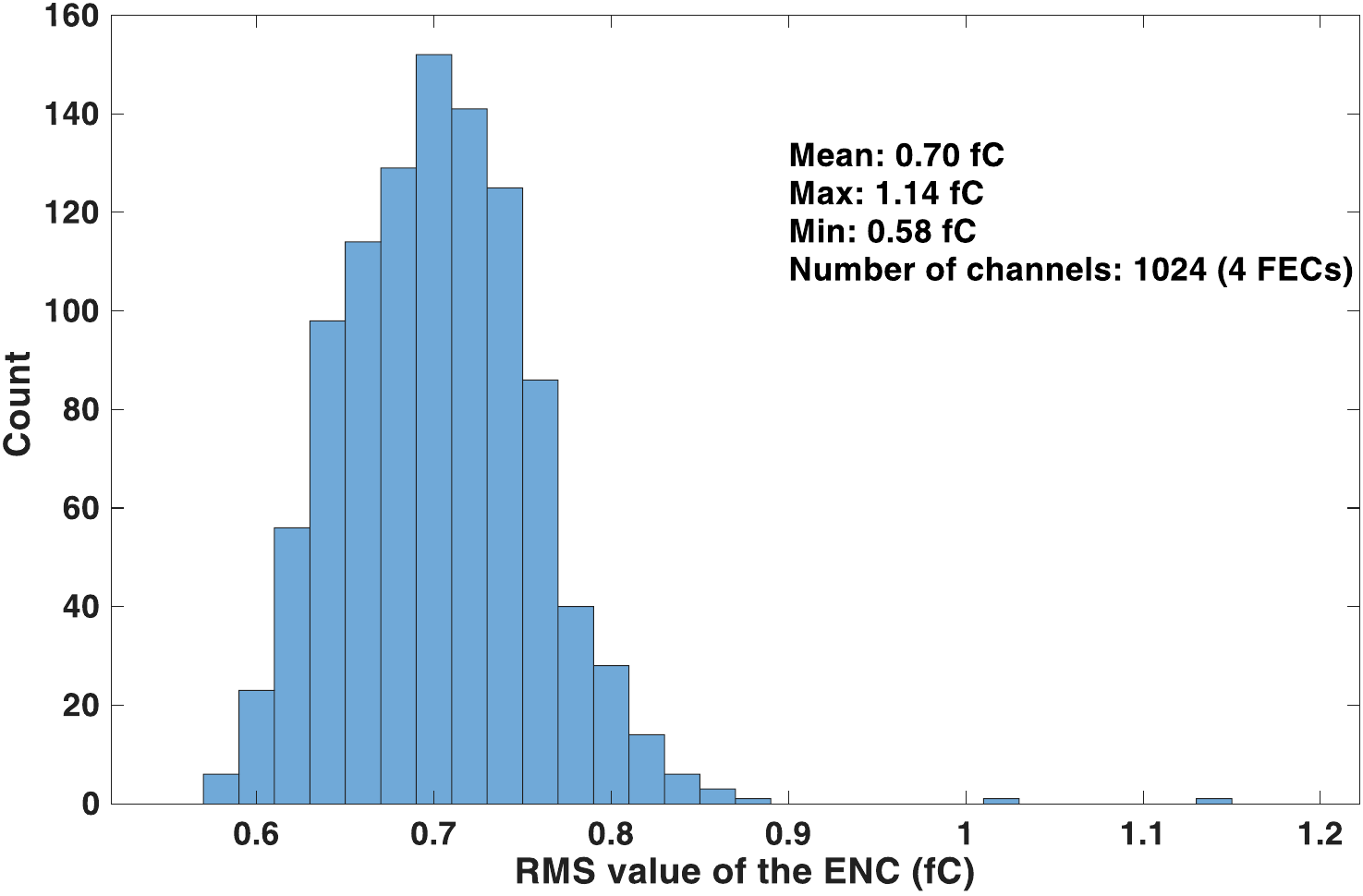}}
\caption{
Distribution of the RMS noise of the FECs connected to the detector.
}
\label{fig8}
\end{figure}

The readout electronics cards were calibrated before installation into the $\upmu$STC-R400. 
In the calibration, a charge signal was generated by applying a step-down voltage through a 1 pF capacitor in series and fed into the calibration pin of the ASICs (In\_cal input).
With the dynamic range set to $\mathrm{120~fC}$ and a peaking time of $\mathrm{1039~ns}$, the charge-to-ADC conversion gain is approximately $\mathrm{29~ADC~units/fC}$.
Under this calibration relationship, the root mean square (RMS) value of the equivalent noise charge (ENC) for all channels connected to the detector via channel multiplexing circuits is shown in Fig.~\ref{fig8}, with 95\% of the FEC channels exhibiting noise less than 0.78 fC.

\subsection{Energy Spectrum and Position Decoding}
The detectors operate in gas-flow mode, with the working gas being a mixture of argon and 7\% carbon dioxide.
In our experiments, the gas flow was set to 25~mL/min and the working voltages of the drift cathode and mesh were set to $-570$~V and $-720$~V, respectively.

\begin{figure}[htbp]
\centerline{\includegraphics[width=0.45\textwidth]{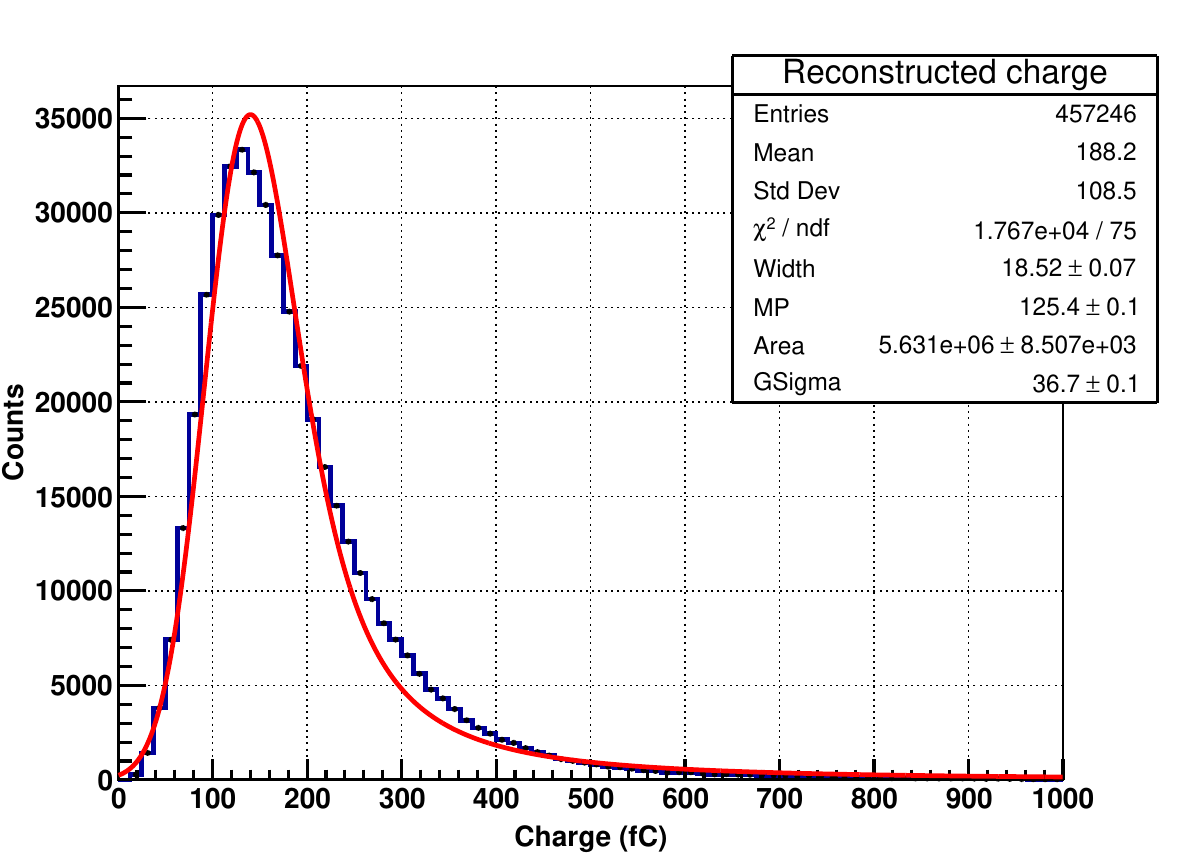}}
\caption{
Energy deposition spectrum of cosmic ray muons measured by the detector, where the blue line with block asterisks represents the measured and reconstructed data, and the red line represents the Convoluted Landau and Gaussian Fit.
}
\label{fig9}
\end{figure}

Each cosmic ray muon deposits energy in the drift area, generating ionization.
This ionization is avalanched in the amplification gap, generating induction signals on the readout strips in both X- and Y-directions.
By summing the charges on the adjacent strips for each valid event, the cosmic ray energy deposition spectrum is obtained and shown in Fig.~\ref{fig9}.
The charge distribution mainly follows the expected Landau distribution. After performing a Landau–Gaussian convolution fit using ROOT, the peak value is approximately $\mathrm{125~fC}$.

\begin{figure}[htbp]
\centerline{\includegraphics[width=0.45\textwidth]{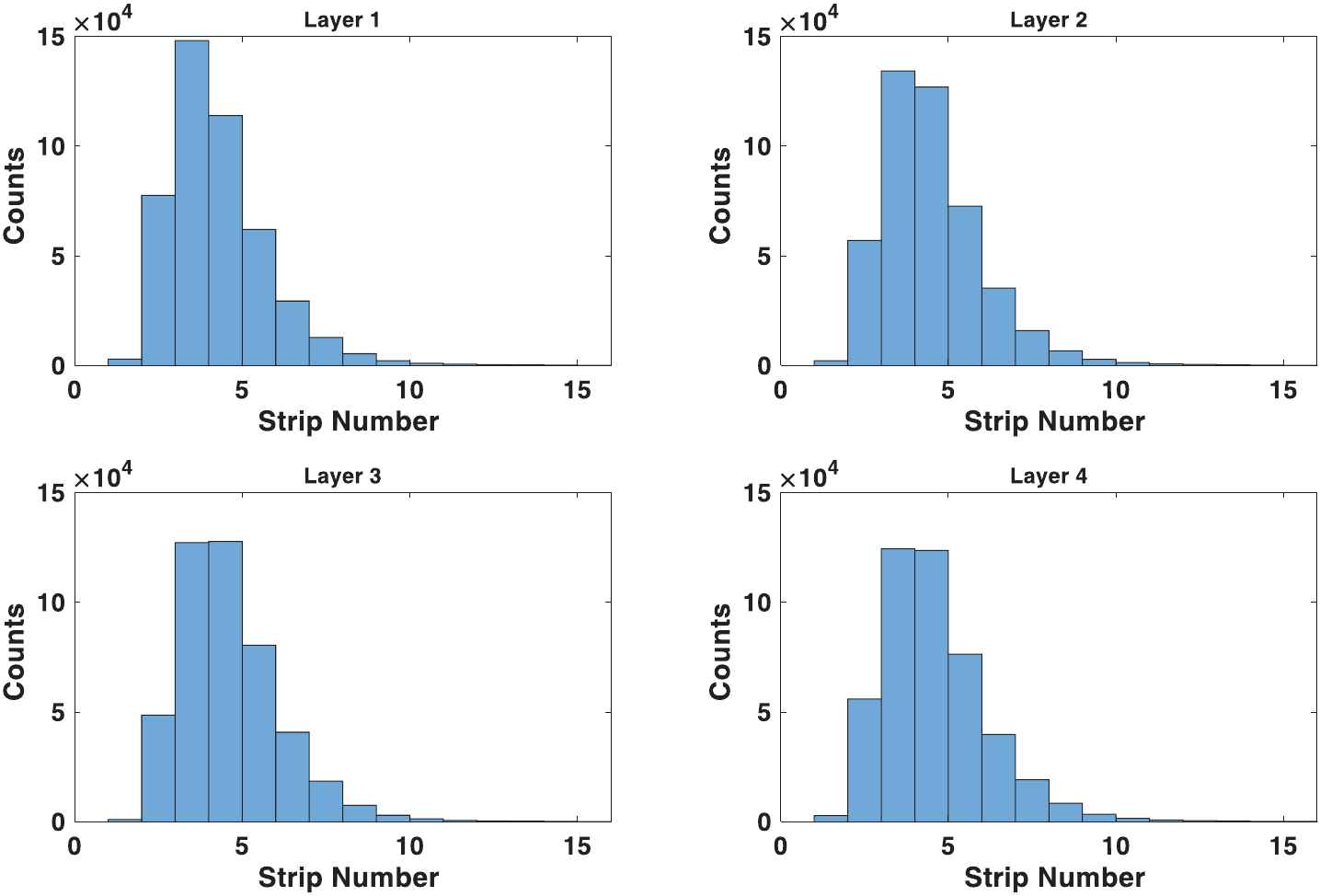}}
\caption{
Distribution of the number of channels exceeding thresholds in each event.
}
\label{fig10}
\end{figure}

Since the strips of the detectors are multiplexed, the hit positions can be decoded when more than one readout electronics channel read out valid signals. 
Figure~\ref{fig10} shows the distribution of the number of channels exceeding thresholds in each event.
This result indicates that the position encoding multiplexing method can achieve nearly lossless channel compression in this muography instrument.

\subsection{Spatial Resolution and Angular Resolution}
\begin{figure}[htbp]
\centerline{\includegraphics[width=0.45\textwidth]{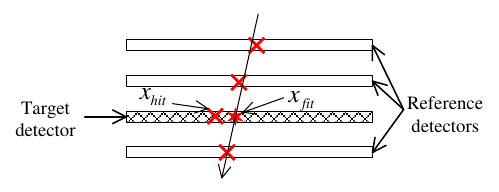}}
\caption{
Illustration of the method for calculating spatial resolution. The actual hit positions are represented by red crosses, while the fit hit position is indicated by a red hexagram.
}
\label{fig11}
\end{figure}

\begin{figure}[htbp]
\centerline{\includegraphics[width=0.45\textwidth]{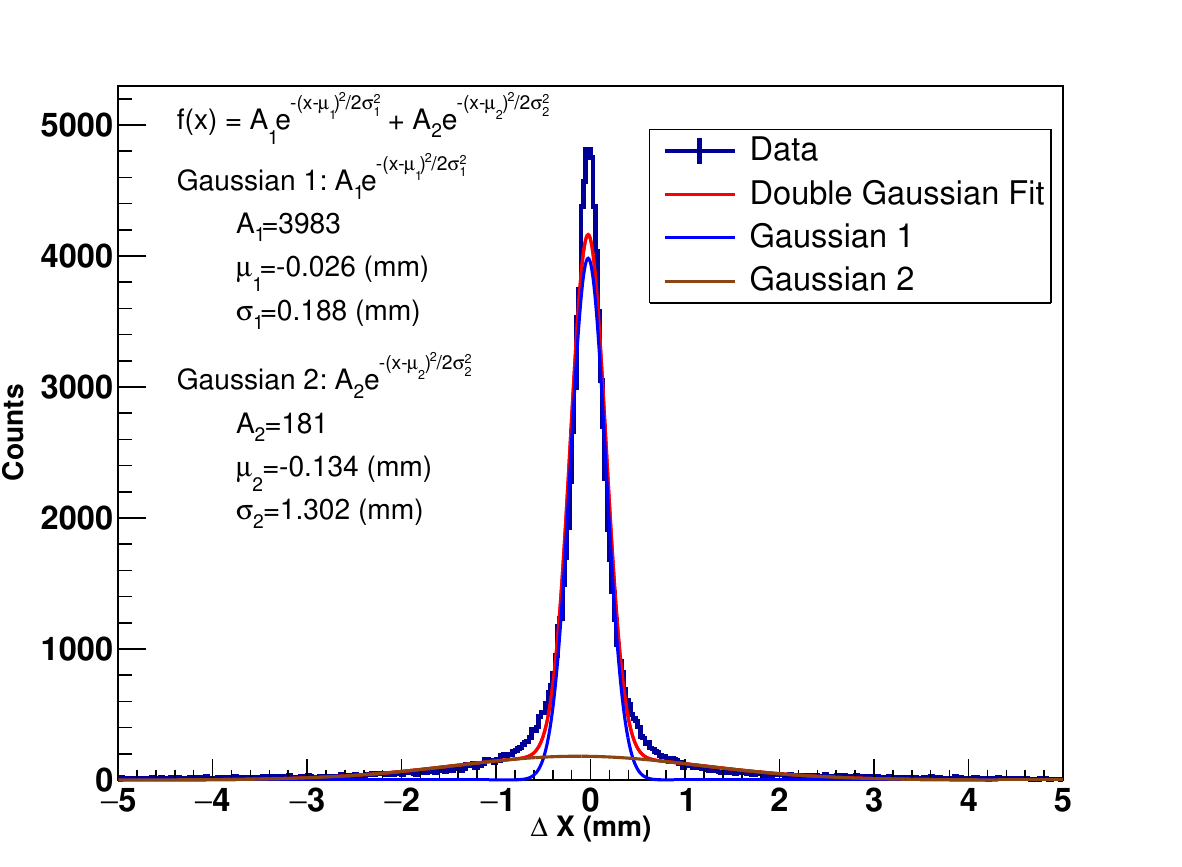}}
\caption{
Deviation distributions of the Micromegas in the X direction with an incident angle range of $0^\circ$ to $5^\circ$. 
}
\label{fig12}
\end{figure}

The principle of calculating the spatial resolution of the instrument is shown in Fig. \ref{fig11}. 
One of the layers is designated as the target layer, and the hit positions from the other three layers are used to fit a reference track.
The fitting result can be expressed as $x_{fit}=k_x\times z_t + b_x$, 
where $k_x$ and $b_x$ are the slope and intercept of the fitting track, respectively, and $z_t$ is the z coordinate of the target layer.
After detector alignment to correct the installation offsets and rotations, the deviation $\Delta x =x_{hit} - x_{fit}$ can be calculated, where $x_{hit}$ is the hit position measured by the target detector. 
By statistically analyzing a large number of muon tracks, the distribution of $\Delta x$ can be obtained and fitted with a double Gaussian function.
Figure \ref{fig12} shows the fitting result of the $\Delta x$ in the X direction of the inner layer detector with an incident angle range of $0^\circ$ to $5^\circ$.
The narrow component of the Gaussian distribution (blue line) has a standard deviation of $0.188 ~\mathrm{mm}$.
The spatial resolution of the instrument is defined as the standard deviation of the $\Delta x$ distribution.
This resolution can be conceptually divided into two contributing factors: the uncertainty from trajectory fitting ($\sigma(x_\text{fit})$) and the intrinsic residual of the detector under test ($\sigma(x_\text{hit})$). 
In muography applications, both these components contribute to the overall precision of muon track measurement, and their combined effect represents the spatial resolution of the instrument.

Regarding angular resolution, the muon track is reconstructed by fitting the hit points from four layers. 
The track can then be represented by a direction vector $\vec{D}(k_x, k_y, 1)$ and a point $P(x_0, y_0, 0)$, where $k_x$ and $k_y$ are calculated as:

\begin{equation}
\begin{aligned}
     k_x = \frac{\sum_{j=1}^{n}{(z_j-\bar{z})x_j}}{\sum_{1}^{n}(z_j-\bar{z})^2}~,~
     k_y = \frac{\sum_{j=1}^{n}{(z_j-\bar{z})y_j}}{\sum_{1}^{n}(z_j-\bar{z})^2}
     \end{aligned}
     \label{eq_kxky}
\end{equation}
where $n$ is the number of fitted detectors and ($x_i$, $y_i$, $z_i$) are the hit points for each detector. Note that the $z_i$ coordinates are fixed after the instrument's construction; for our four layers, these are $\mathrm{0~mm, 167~mm, 395~mm, and~ 587~mm}$.

Thus, the reconstructed angles are given by: $\theta = \arctan\sqrt{k_x^2+k_y^2}$ and $\phi=\arctan(k_y/k_x)$, where $\theta$ and $\phi$ are the zenith and azimuthal angles, respectively. 
The angular resolution of $\theta$ can thus be written as:

\begin{equation}
     \Delta \theta=\frac{1}{1+k_x^2+k_y^2}\times \sqrt{\left(\Delta k_x^2+\Delta k_y^2\right)}
\end{equation}
where $\Delta k_x$ and $\Delta k_y$ can be calculated from Eq.(\ref{eq_kxky}). 
Assuming all detectors have similar performance, $\Delta k_x$ can be approximated as $\Delta k_x=\Delta x/\sqrt{\sum_1^n(z_j-\bar{z})^2}$.
Given that $k_x$ and $k_y$ can be minimal (approaching zero for near-vertical tracks) and assuming similar resolution along the X and Y dimensions, the overall angular resolution $\Delta \theta$ can be estimated as $\sqrt{2}\Delta k_x$.
For the $\upmu$STC-R400, the angular resolution is approximately $\mathrm{0.6~mrad}$.

\subsection{Muon Radiography Experiments Results}
\subsubsection{Measurement at an Underground Tunnel}
\begin{figure}[htbp]
\centerline{\includegraphics[width=0.45\textwidth]{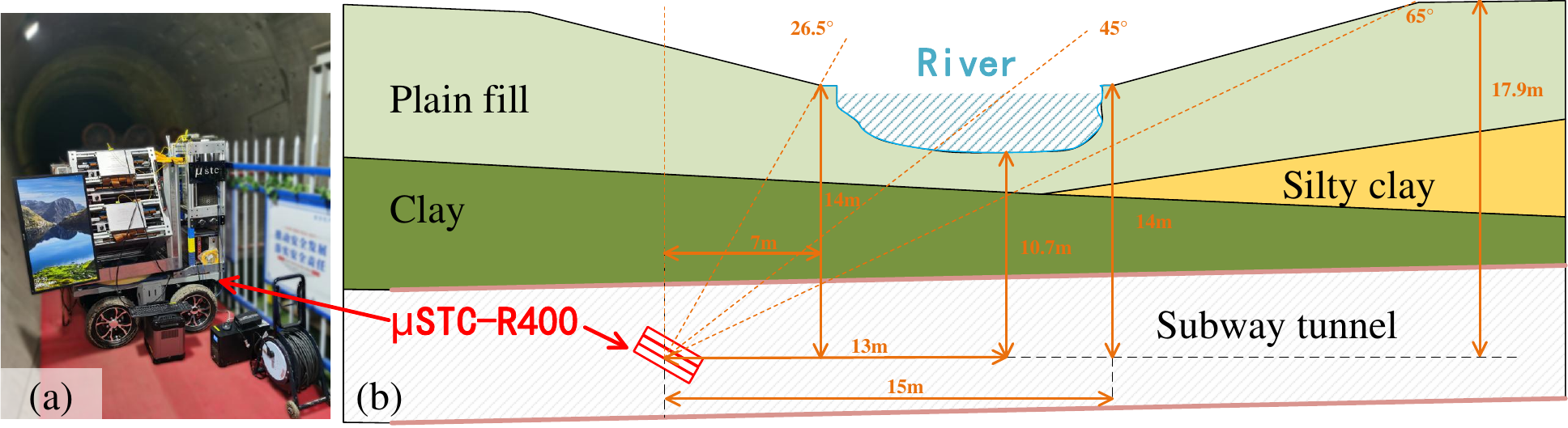}}
\caption{
(a) Photograph of the $\upmu$STC-R400 instrument in the under-construction subway tunnel with a zenith angle of $30^\circ$ and oriented toward the river on the ground surface. The direction of the river is approximately perpendicular to the subway tunnel;
(b) Placement of the detector in the underground tunnel and the cross-section of the stratum from engineering investigation.
}
\label{fig13}
\end{figure}

An experiment was performed at a subway tunnel under construction in the Hefei city.
As shown in Fig.~\ref{fig13}, the $\upmu$STC-R400 was placed in the subway tunnel approximately $18 ~\mathrm{m}$ below the ground surface.
A river on the ground flows above the subway tunnel, approximately perpendicular to it.
According to the engineering investigation, the vertical distance between the riverbed and the bottom of the tunnel is about $11~\mathrm{m}$.
Since the subway site was under construction during the experiment, a UPS module was used to prevent power surges and sudden power loss.
Remote control was implemented via the local WiFi network in the tunnel for monitoring the status of detectors and readout electronics.

\begin{figure}[htbp]
\centerline{\includegraphics[width=0.45\textwidth]{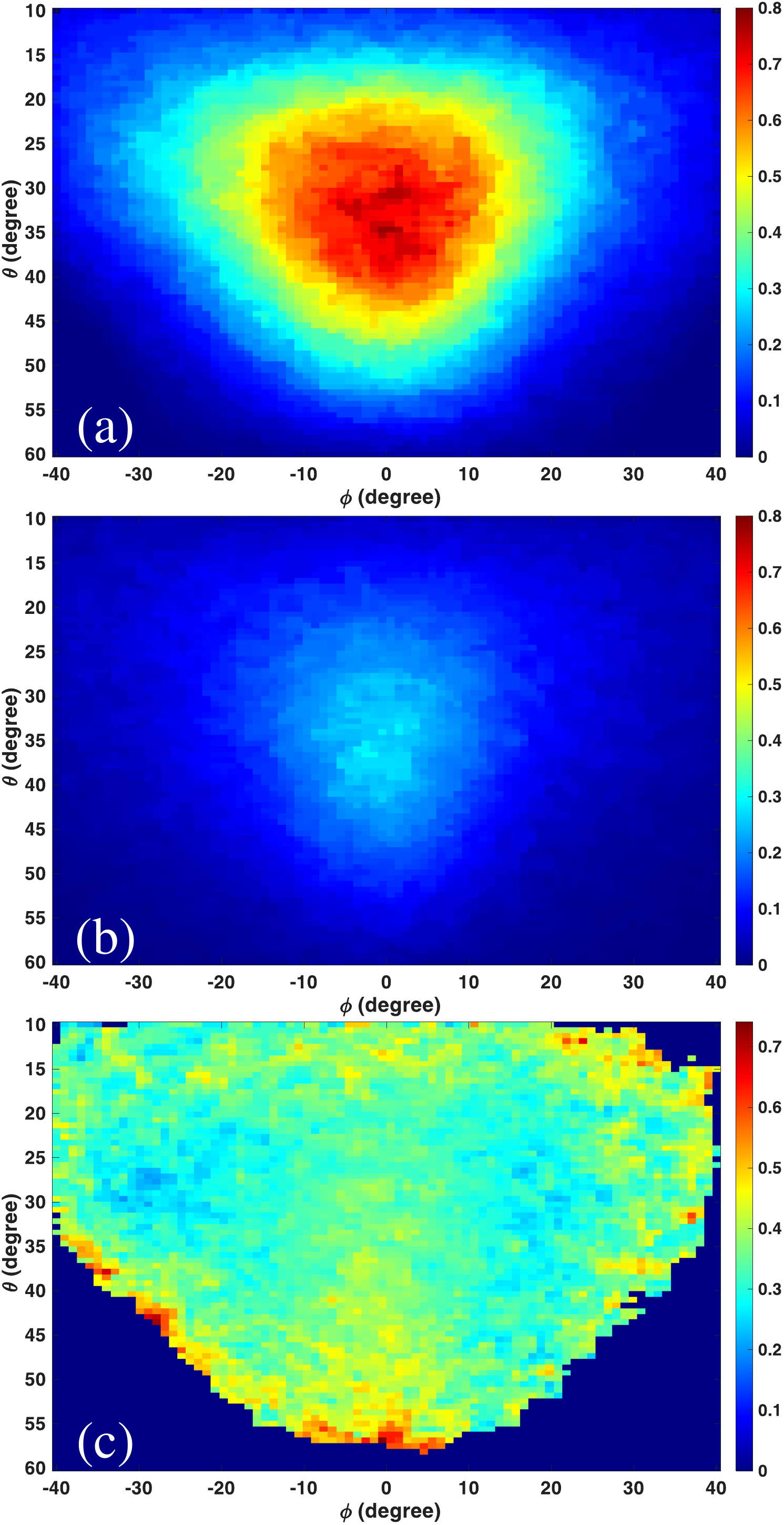}}
\caption{
(a) 2D hit distribution for open-sky measurements;
(b) 2D hit distribution measured in the tunnel.
The color bars in (a) and (b) represent the muon count per 10 minutes.
(c) Transmission map calculated as the ratio of (b) to (a). To suppress statistical noise in low-count regions, pixels with muon counts below 5\% of the maximum are masked (set to zero).
}
\label{fig14}
\end{figure}

\begin{figure}[htbp]
\centerline{\includegraphics[width=0.4\textwidth]{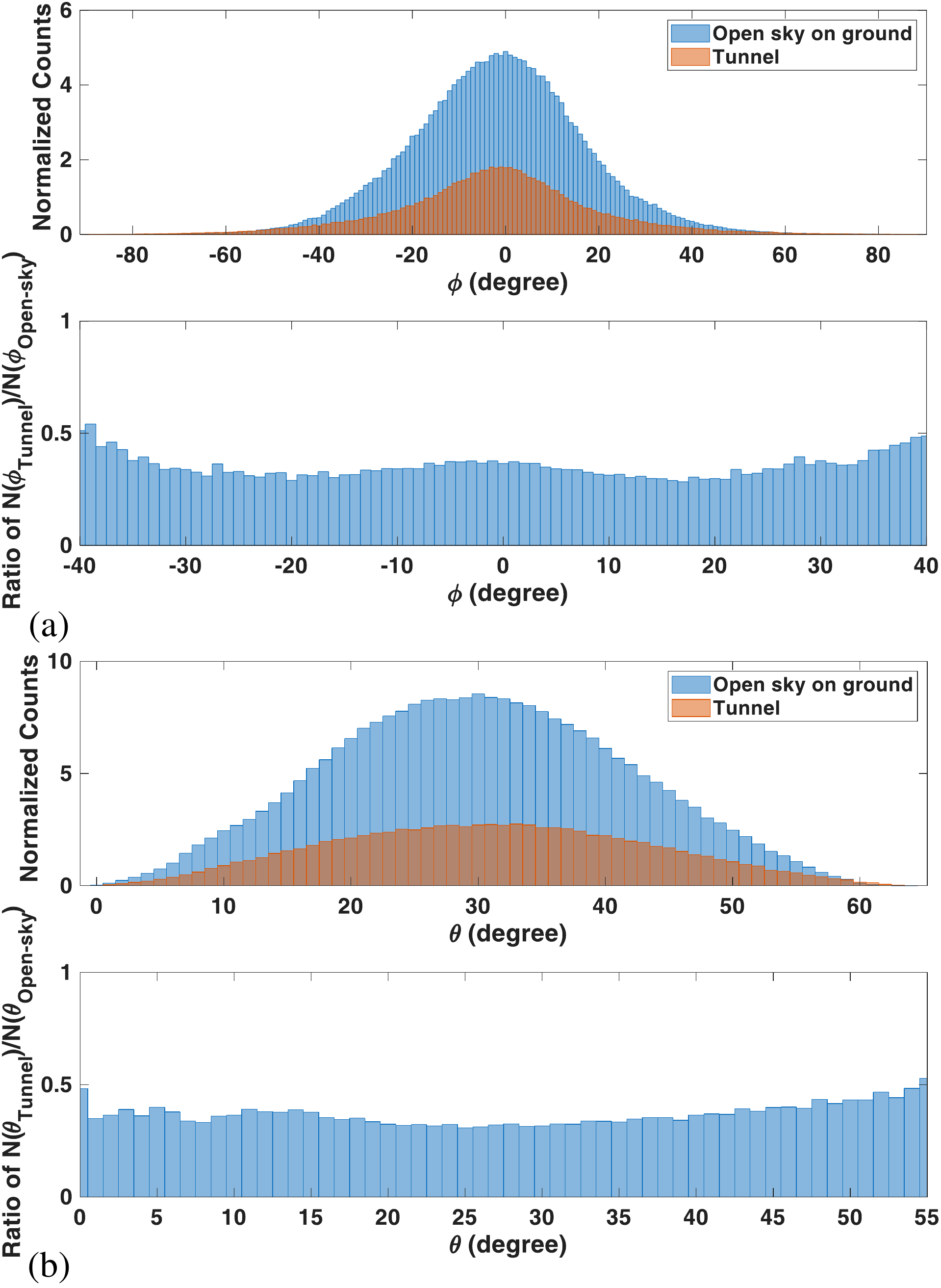}}
\caption{
(a) Distribution of the $\theta$ angle and the ratio $\theta(N_{\text{tunnel}})/\theta(N_{\text{open-sky}})$;
(b) Distribution of the $\phi$ angle and the ratio $\phi(N_{\text{tunnel}})/\phi(N_{\text{open-sky}})$.
}
\label{fig15}
\end{figure}

The measurement in the tunnel lasted approximately 300 hours, while an open-sky muon flux measurement conducted at ground level on the USTC campus lasted about 266 hours.
After performing position alignment to correct for offset and rotation errors,
muon tracks were reconstructed where all 4 layers of detectors were fired and the RMSE (Root Mean Square Error) of the fitted tracks was less than $2~\mathrm{mm}$.
Figure~\ref{fig14} (a) and (b) show the 2D hit distributions measured on the ground (the open sky) and in the tunnel, respectively, with the muon counts normalized to 10 minutes. 
Figure~\ref{fig14} (c) presents the transmission map for the subway tunnel measurement.
Figure~\ref{fig15} shows the distributions of the $\theta$ and $\phi$ angles for open-sky and tunnel measurements (counts normalized to 10 minutes).
Results show that the muon flux is attenuated by the soil above the tunnel, while the influence of the river is not significant.
This can be attributed to the limited density difference between the river water and the surrounding overburden soil.
Nonetheless, the $\upmu$STC-R400 has demonstrated its robustness by operating successfully in the subway tunnel environment where vapor and dust were present during the test.

\subsubsection{Measurement at a Mountain}
\begin{figure}[htbp]
\centerline{\includegraphics[width=0.45\textwidth]{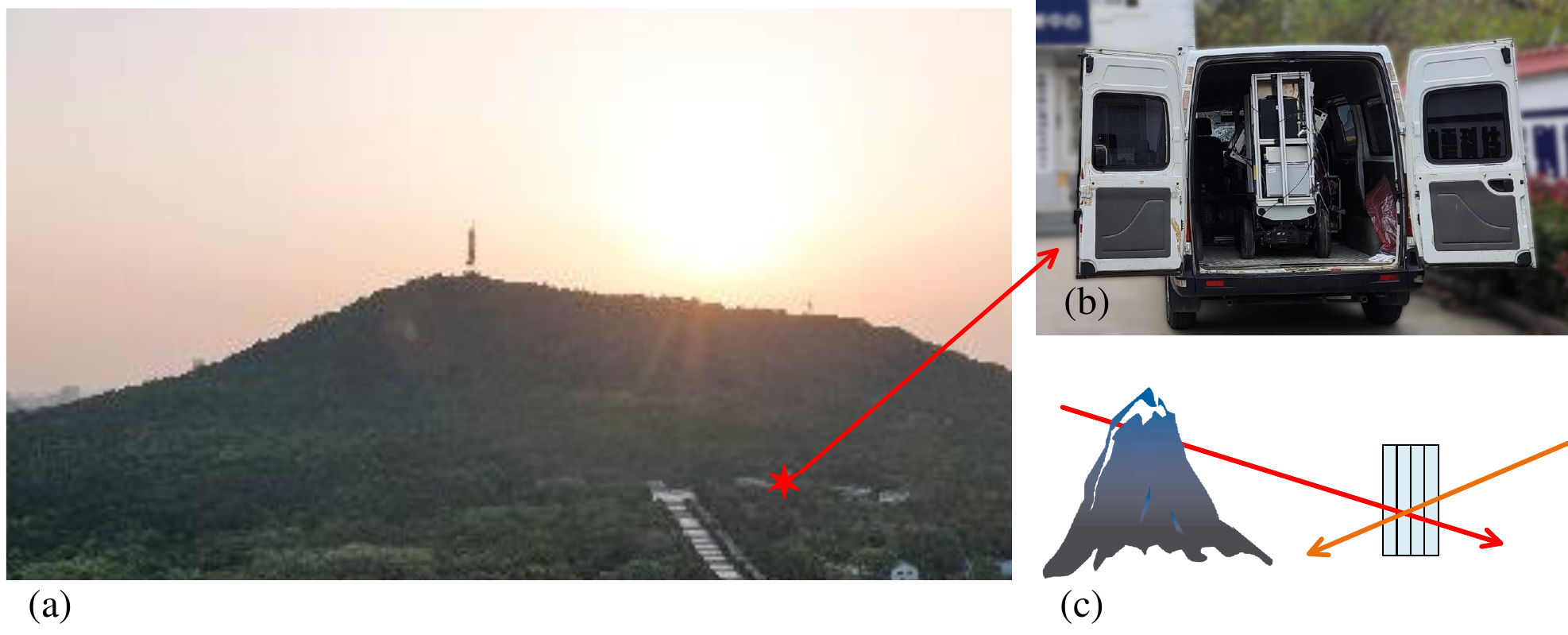}}
\caption{
(a) View of Dashu Mountain, with the $\upmu$STC-R400 placed at the foot of the mountain (approximately at the location marked by the hexagram (with GPS location at 117.18°E 31.84°N 74.7 m));
(b) Photograph of the instrument placed in a microbus;
(c) Diagram of the detector placement, where the normal of the detector is parallel to the ground. 
}
\label{fig16}
\end{figure}

To validate the versatility of this muography instrument in outdoor environments, an experiment was performed at Dashu Mountain in the Shushan Forest Park located in the urban area of Hefei city.
As shown in Fig.~\ref{fig16}(a) and (b), the $\upmu$STC-R400 was placed in a microbus positioned near the foot of the mountain.
The horizontal distance between the instrument and the top of Dashu Mountain is approximately 1 km, with a vertical distance of about 200 m.
To measure both the open-sky muon flux and the flux through the mountain at the same time, the detectors were rotated so that their normal was parallel to the ground, with the center approximately pointing toward the mountain peak.
As shown in Fig.~\ref{fig16}(c), the detectors could simultaneously record muons from both the mountain direction and the open sky.
This configuration ensured that fluctuations in muon flux and detector gain remained equal for both the target measurement and the background measurement.

\begin{figure}[htbp]
\centerline{\includegraphics[width=0.45\textwidth]{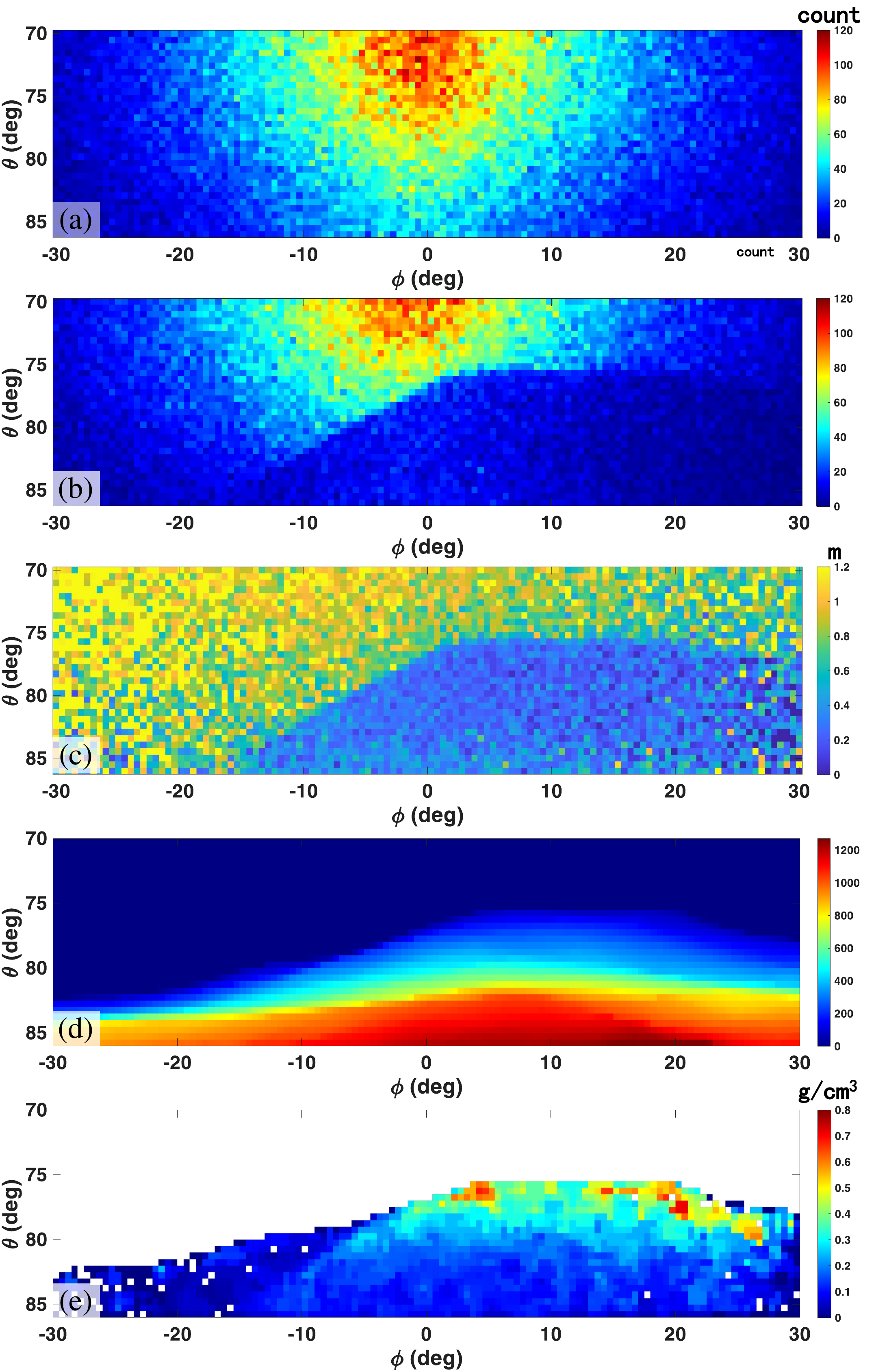}}
\caption{
(a) 2D hit distribution from the open-sky direction;
(b) 2D hit distribution from the mountain direction;
(c) Transmission ratio calculated as $N_{\text{mountain}}/N_{\text{open-sky}}$ with values larger than 1.2 set to 1.2;
(d) Traversal length from the experiment site perspective;
(e) Reconstructed density using a constant minimum transmission energy of $\mathrm{2.5~MeV\cdot g^{-1}\cdot cm^2}\times X$.
}
\label{fig17}
\end{figure}
The test continued for approximately 37 days, during which a heavy snowfall occurred in Hefei, yet the instrument maintained proper functioning.
The 2D distribution of muons from the open-sky direction and from the mountain direction is shown in Fig.~\ref{fig17}(a) and (b), where the 'shadow' caused by the mountain's absorption is clearly visible.
Figure~\ref{fig17}(c) shows the transmission ratio, where values larger than 1.2 were set to 1.2 to provide a clear color gradation.
The shape of Dashu Mountain was obtained through drone mapping with an accuracy of $10~\mathrm{m}$, and Fig.~\ref{fig17}(d) shows the traversal length at the experiment site.

With the known transmission rate $T(\theta,~\phi)$, the traversal length $L(\theta,~\phi)$, and according to Eq.(\ref{eq_opacity}),
the average density of the object can be calculated as $\bar{\rho}(\theta,~\phi) = X( \left.\theta,~\phi\right| E_{\text{min}}) / L(\theta,~\phi)$,
where $E_{\text{min}}$ is the minimum energy required to traverse the objects.
The reconstructed density is shown in Fig.~\ref{fig17}(e) using parameters from the EcoMug library~\cite{paganoEcoMugEfficientCOsmic2021} and from this work~\cite{marteauDIAPHANEMuonTomography2017},
with the mountain body treated as rock with $E_{\text{min}}=\mathrm{2.5~MeV\cdot g^{-1}\cdot cm^2}\times X$.
The reconstructed density values are lower than expected empirical values. 
This discrepancy may be attributed to uncertainties in the determination of $E_{\text{min}}$ at large incident angles, as well as the $\mathrm{10~m}$ accuracy of the traversal length measurement obtained from drone mapping.

\section{Discussion and Conclusions}
In this study, a high spatial resolution muon radiography instrument named $\upmu$STC-R400 was designed and implemented.
Four layers of detectors with a sensitive area of $\mathrm{400~mm\times 400~mm}$ were used, requiring a total of 8000 strips to be read out.
A channel multiplexing method was applied to reduce the large number of readout channels. 
Together with the scalable readout system and the self-trigger mode, this instrument could be constructed in a relatively compact size.
Test results show that the RMS of the ENC is approximately 0.8 fC with a dynamic range of $0\sim120~\mathrm{fC}$, which adequately covers the signal range of the Micromegas detector.
The spatial resolution of the detectors with encoding readout is approximately $\mathrm{190~\upmu m}$, which provides sufficient angular resolution for muon radiography. 
The long-term stability and environmental suitability were tested during different types of experiments in harsh environments, demonstrating that with proper design and sealing, the Micromegas detectors can operate effectively in muography experiments outside the laboratory.
The instrument presented in this article indicates that Micromegas detectors with the thermal bonding method can be a promising solution for building high spatial resolution muography facilities, while the channel multiplexing readout circuit significantly reduces system complexity.

Current research and development efforts are ongoing toward further enhancement of the instrument. 
In the future, more advanced algorithms can be developed based on the high-resolution muon track data. Specifically, the energy spectrum model at large incident angles requires further investigation.
The current angular resolution significantly exceeds the requirements for muon transmission radiography. This suggests that for future high spatial resolution detector applications, the distance between detectors can be shortened. A reduced inter-detector distance would enable the instrument to achieve a larger field of view, thereby increasing the muon rate.

\begin{acknowledgments}
This work was financially supported in part by the National Natural Science Foundation of China (NSFC) for Distinguished Young Scholars under Grant No.~12025504, in part by the National Natural Science Foundation of China (NSFC) for Young Scientists under Grant No.~12205297, and in part by the University of Science and Technology of China.

The authors would like to express their sincere gratitude to Yong Zhou, Sicheng Wen, Han Han, Xing Xu, Haibin Fei, Bo Wang, Mujun Li, Jiale Gao, and Xiaozhao Cao from Jianwei Scientific Instruments (Anhui) Technology Co., Ltd. for their help in detector manufacturing, instrument construction and experiments deployment.

The authors also extend their appreciation to China Construction Eighth Engineering Division CORP., LTD. for their assistance with the experimental site in the subway tunnel.

The authors are grateful to Haigang Zheng, Hongbo Sun, and Xiao Liang from Anhui Earthquake Agency for their support at the Dashu Mountain experimental site.

\end{acknowledgments}

\section*{Data Availability Statement}
The data that support the findings of this study are available from the first author and the corresponding author upon reasonable request.

\section*{References}
\bibliography{mustc_400_3rd_arxiv.bib}

\end{document}